\documentclass[11pt,nofootinbib,aps,preprint]{revtex4-1}

\usepackage{amsmath,mathrsfs}
\usepackage{graphicx,epsfig}
\usepackage{latexsym,amssymb}
\usepackage{epsf}
\usepackage{ifpdf,natbib,lineno}

\usepackage{xcolor}

\begin{document}

\title{{\bf  The role of the massless phantom term in  the stability of a non-topological soliton solution }}
\author{ M. Mohammadi} \address{Physics Department, Persian Gulf University, Bushehr 75169, Iran.}
\email{physmohammadi@pgu.ac.ir}


\begin{abstract}

We intend to introduce classically a special Lagrangian density in such a way that, firstly, it leads to a special non-topological  solitary wave solution, secondly,  the stability of that is guaranteed properly, and thirdly, its  dominant dynamical equations  reduce  to  the   standard nonlinear Klein-Gordon equations. For these purposes, we have to consider a new term in the Lagrangian density, whose role is like a massless phantom that surrounds the special  solitary wave solution  and resists  any change in its internal structure.

\end{abstract}

\maketitle


\section{Introduction}

In many of known physical models in the quantum field theory, dynamical field equations are introduced as standard (nonlinear) Klein-Gordon or (nonlinear) Klein-Gordon-like equations. Each of these equations with a set of particular constants just is used for a special  type of fundamental particles with specific characteristics. For example, the components of Dirac equation (as some Klein-Gordon  equations) were initially introduced to describe  the quantum behavior of electrons and positrons. The Dirac equation, with different constants, is used for neutrinos and muons separately. Also, for the known Higgs particles, the  dominant dynamical  equation is a complex nonlinear Klein-Gordon (KG) equation. Nevertheless, the truth and the nature of the particle itself remain unknown in the standard quantum field theory. In other words, the quantum field theory is just a mathematical structure that provides a correct probabilistic   relationship between initial conditions and the output results. Many of the properties of the fundamental particles in the quantum field theory (such as mass and charge) are based solely on the results that obtained in the laboratory and  are used manually to obtain the appropriate dynamical equations in this standard and successful theory. In fact, the quantum field theory, despite many of its significant successes in predicting the probability behavior of particles, in many other respects, can not explain many of other questions. For example, why should the mass and charge of a fundamental particle, such as an electron, be specific numbers? Or, why is there just   a single Planck constant $\hbar$ for the all particles of the nature? These  are some of the questions that have  not been yet explained     properly.

Classically, for any  arbitrary  system of the fields, there are  some PDEs which are called  the   dynamical wave equations. In general, theses PDEs have infinite solutions but   a special solution with minimum rest energy is an important one, the so-called  a soliton solution \cite{rajarama,Das,lamb,Drazin,TS}. A soliton solution in many respects is in accordance with our classical sense of the particle concept, i.e.  a prominent stable profile  of the field in the space whose behavior in the collisions is fully in line with well-known standard theories. For example, for the real nonlinear KG systems with kink (antikink) solutions, many of the expected properties, that we expect to satisfy for any  real classical relativistic particle, are satisfied properly \cite{rajarama,TS,phi41,phi42,phi43,phi44,phi45,OV,GH,MM1,MR,DSG1,DSG2,DSG3,MM2,Kink1,Kink2,Kink3,Kink4,Kink5,Kink6,Kink7,Kink8,waz}. In fact, the theory of the classical fields with soliton solutions is a hopeful point for some researchers to answer the unanswered questions that remain in the quantum field theory. The stability condition for soliton solutions has forced many researchers to look for models that result  topological solutions \cite{rajarama,TS,SKrme,SKrme2,toft,pol,Vash,Mahzon}. Basically, the topological property causes the stability of the soliton solution  to be automatically guaranteed. Instead, among the models with non-topological solutions \cite{waz,waz2,Vak3,Vak4,Vak5,Vak6,Lee3,Scoleman,R1,R2,R3,R4,R5,R6,R7,R8,Riazi2,MM3,MM4}, no model has been yet introduced that can lead to a soliton solution with minimum  energy. The importance of the non-topological solutions is that it is easy to imagine a multi-particle solution just by adding them when they are far enough from each other. In the case of topological solutions to have a multi-particle solution, the situation is usually very complicated and sometimes impossible. Note that, the new  model
in this paper is introduced   in 1 + 1 dimension just for simplicity, but it can be extended to 3 + 1 dimensions
in a similar way.

In this paper, inspire by  what we have learned in the quantum field theory, we try to show how it is possible to introduce a classical model of the relativistic fields in such a way that it leads to  a special  non-topological soliton solution with a special type of the standard nonlinear KG equations as its dominant dynamical equations.  We show that in order to achieve these demands, we have to add   a new  term  to  a standard nonlinear KG  Lagrangian density  which behaves like a massless phantom that  surrounds its  special solitary wave   solution  and  obstructs any arbitrary  change in the internal structure of that. In other words, we want it acts like a stability catalyzer and its role   be  hidden  when the special solitary wave solution is free and non-deformed.  This new special classical relativistic  field   model  is an example of  the extended nonlinear  KG field systems that are introduced in this paper generally.

The organization of this paper is as follows: In Section \ref{s2}, we introduce the standard  and the extended (nonlinear) KG systems for scalar fields. In Section \ref{s3}, in general, we will  introduce an extended  nonlinear  KG system with a special non-topological soliton solution. In Section \ref{s5}, the stability  of the special soliton solution is considered under the small variations. The last section is devoted to  summary and conclusions.

\section{Extended nonlinear Klein-Gordon systems}\label{s2}

In the  standard  relativistic (classical and quantum) field   theory, the standard forms  of the Lagrangian  densities for the scalar fields are   expressed as follows:
 \begin{equation} \label{asd}
{\cal L}= \sum_{j\geqslant i}^{N}\sum_{i=1}^{N} \alpha_{ij}(\phi_{1},\cdots,\phi_{N})S_{ij}-V(\phi_{1},\cdots,\phi_{N}),
 \end{equation}
here $\phi_{i} $'s  are $ N $ independent scalar fields, coefficients $\alpha_{ij}$'s  and potential term $V$ all are functions of the fields, scalars $S_{ij}=S_{ji}=\partial_{\mu}\phi_{i}\partial^{\mu}\phi_{j}$ are named kinetic scalar terms.  In fact, such kind of Lagrangian densities  represent  the standard (nonlinear) KG systems.  In other words, the formulas of the Lagrangian density of the  standard  (nonlinear)  KG systems  are linearly expanded in terms of the kinetic scalar  terms $ S_{ij} $. Note that, the various linear combinations of $ S_{ij} $'s, in accordance with the standard (\ref{asd}), are ones that result a real Lagrangian density.
However, according to the standard (\ref{asd}), depending on the arbitrary choices of the coefficients $ \alpha_{ij} $'s  and potential  term  $ V $, it  can be possible to introduce  infinite (nonlinear) KG system. For example, if we deal with a complex scalar field $ \phi $, then $ \phi_{1} = \phi $, $ \phi_{2} = \phi^{*} $ and the allowed kinetic scalar  terms are  $ S_{11} = \partial_\mu \phi \partial^\mu\phi $, $ S_{22} = \partial_\mu\phi^* \partial^\mu \phi^* $ and $ S_{12} = \partial_\mu \phi \partial^\mu \phi^* $.  Accordingly, for a complex scalar field $ \phi $, the well known systems with the non-topological solitary wave  solutions  (Q-balls) were  introduced  as follows \cite{waz,waz2,Vak3,Vak4,Vak5,Vak6,Lee3,Scoleman,R1,R2,R3,R4,R5,R6,R7,R8,Riazi2,MM3,MM4}:
\begin{equation} \label{exam}
{\cal L}= \partial_\mu \phi^*
\partial^\mu \phi -V(R),
 \end{equation}
where $\alpha_{11}=\alpha_{22}=0$, $\alpha_{12}=1$ and  $R=|\phi |=\sqrt{\phi \phi^{*}}$.
Q-balls, unlike kinks and anti-kinks, are non-topological solitary wave solutions which are not energetically stable. In general, it was shown that Q-balls have the minimum rest energy among the other solutions with the same electrical charge, but it is not a sufficient condition for the stability \cite{Vak6,Scoleman}.

In an equivalent representation, instead of a  complex scalar field $ \phi $, using the polar representation of the fields
 \begin{equation} \label{polar}
 \phi(x,t)= R(x,t)\exp[i\theta(x,t)],
\end{equation}
leads to a new form of the systems with the Q-ball  solutions, i.e.
\begin{equation} \label{cnk}
{\cal L}=(\partial^\mu R\partial_\mu R) +R^{2}(\partial^\mu\theta\partial_\mu\theta)-V(R),
\end{equation}
where $\phi_{1}=R$, $\phi_{2}=\theta$, $S_{11}=\partial_{\mu}R\partial^{\mu}R$, $S_{22}=\partial_{\mu}\theta\partial^{\mu}\theta$, $S_{12}=\partial_{\mu}R\partial^{\mu}\theta$, $\alpha_{11}=1$, $\alpha_{22}=R^2$  and $\alpha_{12}=0$.  Therefore, a special relativistic field  system can be introduced via  many different, but equivalent, representation. For each of these equivalent representation, the coefficients $ \alpha_{ij} $'s  and the kinetic  scalar terms  $ S_{ij} $'s would be  different.

If the relativistic Lagrangian densities  of the scalar fields are not  linear  combination of  $ S_{ij} $'s, we can call  them  \emph{extended nonlinear KG} systems.
Namely, for a real scalar field $ \phi_{1} = \varphi $ with a single kinetic  scalar term  $ S_{11} = S = \partial _ {\mu} \varphi \partial ^{\mu} \varphi $, the following Lagrangian densities are two  examples of the extended  nonlinear KG systems:
\begin{eqnarray} \label{kex}
&&{\cal L}=\varphi S+S^2-V(\varphi), \\ \label{kex2}&&
{\cal L}=[S-V(\varphi)]^2.
\end{eqnarray}
Moreover, for a complex field $ \phi $ in the polar representation, i.e.  $ \phi_{1} = R $ and $ \phi_{2} = \theta $, the following Lagrangian densities are again two  examples of the extended nonlinear  KG systems:
\begin{eqnarray} \label{ddd}
&&{\cal L}=(\partial^\mu R\partial_\mu R) +R^2(\partial^\mu\theta\partial_\mu\theta)^3=({\cal S}_{12}) +R^2({\cal S}_{22})^3, \\ \label{ddd2}&&
{\cal L}=(\partial^\mu R\partial_\mu R) (\partial^\mu\theta\partial_\mu\theta)-V(R)=({\cal S}_{11}) ({\cal S}_{22})-V(R)).
\end{eqnarray}
 The extended nonlinear  KG systems, due to their nonlinear dependence on the kinetic scalars  $S_{ij}$, undoubtedly end up with nonlinear dynamical equations of  motion which are too complicated and are unlikely to be used until now in well-known physical models.
 Note that, the pervious Eqs.~(\ref{kex})-(\ref{ddd2}) are just some arbitrary examples of the extended nonlinear KG systems  and there is  not any  other meaning, i.e. they should    not be  considered as some extended field systems with some special solutions or some thing else. Of course, in general,  such Lagrangian densities  can be called nonstandard Lagrangian (NSL) densities  too \cite{SH,ZZ,Rami0,Rami,Rami2}. There are many works which were dealing with such systems among which one can mention the  works of Riazi and his colleagues \cite{Vash,Mahzon}  and El-Nabulsi \cite{Rami,Rami2}. However, in the next section we will introduce a special extended nonlinear  KG system which leads to a single non-topological  soliton solution.

\section{Introducing an extended nonlinear  Klien-Gordon system with a special  non-topological soliton solution}\label{s3}

Here, at first step,  we are going to introduce a standard  complex non-linear KG system with a special  Q-ball solution. For example, if we chose  a special potential
\begin{equation} \label{SSS}
V(R)=-R^6+R^4+100R^2,
\end{equation}
 for Lagrangian density (\ref{cnk}), then the related equations of motion would be
\begin{eqnarray} \label{W1}
&&\Box R-R(\partial^\mu\theta\partial_\mu\theta)=-\frac{1}{2}\frac{dV}{dR}, \\ \label{W2}&&
\partial_{\mu}(R^2\partial^{\mu}\theta)=2R(\partial_{\mu}R\partial^{\mu}\theta)+R^{2}(\partial^\mu\partial_\mu\theta)=0.
\end{eqnarray}
It is easy to show that these dynamical equations have a special Q-ball solution as follows:
 \begin{equation} \label{A0}
\phi=R_{s}(x)e^{i\omega_{s}t}=\dfrac{1}{\sqrt{1+x^2}}e^{i 10 t},
\end{equation}
where  $\omega_{s}=10$ can be called  the rest frequency of this special solution. This solution is at rest, to obtain the moving version of that, if it moves  at velocity  of $ v $, it can be easily accomplished using a relativistic boost:
 \begin{equation} \label{A1}
\phi=\dfrac{1}{\sqrt{1+\gamma^2(x-vt)^2}}e^{i k_{\mu}x^{\mu}},
\end{equation}
where  $\gamma=(1-v^2)^{-1/2}$  and $k^{\mu}\equiv(\omega,k)$, provided
$\omega=\omega_{s}\gamma$ and $k=\omega v$. Note that, in this paper, for simplicity, we take the speed of light equal to one ($c=1$).

The  energy density function belonging  to Lagrangian density (\ref{cnk}) can be easily obtained:
 \begin{eqnarray} \label{A2}
&&\varepsilon(x,t)=\left[\dot{R}^{2}+R'^{2}+R^2 (\dot{\theta}^{2}+\theta'^{2})+V(R)\right],
\end{eqnarray}
where prime and dot are used  to specify the  space and time derivatives  respectively. Since for the non-moving special Q-ball solution (\ref{A0}) $\dot{R_{s}}=\theta_{s}'=0$, therefore the rest  energy of  that would be
\begin{eqnarray} \label{A3}
&&E_{o}=\int_{-\infty}^{+\infty}[R_{s}'^2+\omega_{s}^2R_{s}^2+V(R_{s})]dx.
\end{eqnarray}
It is easy to show that the  non-topological Q-ball solution (\ref{A0})  is not stable under any arbitrary  small deformation. For example, if we fix the phase function $\theta=\omega_{s}t$, according to Eqs.~(\ref{A3}) and (\ref{SSS}), it is easy to show   any
small variation in $R_{s}$ with $\delta R_{s}<0$, yields a small reduction in the related total energy. In other words, the  Q-ball solutions, such as (\ref{A0}), are not energetically stable objects at all.

Our main goal in this paper is that to find a proper additional term ($F$) for the original Lagrangian density (\ref{cnk}) in such a way that, like a stability catalyzer,  guarantees the energetically stability of the  special Q-ball solution (\ref{A0}) and dose not have any role in the dominant dynamical equations  for this special solution, i.e. the dynamical equations  just for the  special Q-ball solution (\ref{A0}) remain the same standard  original equations (\ref{W1}) and (\ref{W2}). In other words, we want to introduce  a new extended nonlinear Lagrangian density as follows
  \begin{equation} \label{LN}
 {\cal L}_{N}= {\cal L}+F=\left[\partial^\mu R\partial_\mu R +R^{2}(\partial^\mu\theta\partial_\mu\theta)-V(R)\right]+ F,
 \end{equation}
in such a way that  the pervious solitary wave function  (\ref{A0}) being  a solution again. Moreover, we expect the new dynamical equations of this new extended system (\ref{LN}) reduce to the same original ones (\ref{W1} and \ref{W2}) just  for the special solitary wave solution (\ref{A0}).   In  general, since the Lagrangian density  (\ref{LN}) must be scalar, so the new unknown term $ F $ can be only a function of the allowed  scalars. Allowed  scalars are the module field  $ R $, the phase field $ \theta $, and the kinetic scalar terms  $\partial_{\mu}R\partial^{\mu}R$, $\partial_{\mu}\theta\partial^{\mu}\theta$ and $\partial_{\mu}R\partial^{\mu}\theta$. To keep the charge conservation law again, $ F $ should not depend explicitly on $ \theta $. However, the new form  of  the dynamical equations  of the new Lagrangian density (\ref{LN}) are obtained as follows:
\begin{eqnarray} \label{geq}
&&\Box R-R(\partial^\mu\theta\partial_\mu\theta)+\frac{1}{2}\frac{dV}{dR}+\frac{1}{2}\left[ \frac{\partial}{\partial x^{\mu}}\left(\frac{\partial F }{\partial (\partial_{\mu}R)}\right)-\left(\frac{\partial F}{\partial R}\right)  \right]=0\\ \label{geq2}&&  \partial_{\mu}(R^2\partial^{\mu}\theta)+\frac{1}{2} \left[ \frac{\partial}{\partial x^{\mu}}\left(\frac{\partial F }{\partial (\partial_{\mu}\theta)}\right)  \right]=0.
\end{eqnarray}
Based on the well-known relations in the standard classical field  theory, it is easy to find the energy density function  of the system (\ref{LN}):
 \begin{eqnarray} \label{e1}
&&\varepsilon(x,t)=\varepsilon_{o}+\varepsilon_{F}=\left[\dot{R}^{2}+R'^{2}+R^2(\dot{\theta}^{2}+\theta'^{2})+V(R)\right]+\left[\dot{R}\frac{\partial F}{\partial \dot{R}}+\dot{\theta}\frac{\partial F}{\partial \dot{\theta}}\right]
\end{eqnarray}
Where $ \varepsilon_{o} $ corresponds to the original   Lagrangian density (\ref{cnk}) and $ \varepsilon_{F} $ is related to the unknown new added term  $ F $.  As we have mentioned, for the  special solution (\ref{A0}), we expect the equations of motion (\ref{geq}) and (\ref{geq2})  reduce   to the same original standard nonlinear KG equations  (\ref{W1}) and (\ref{W2}) respectively. In other words, for the new system (\ref{LN}), we expect the localized wave function (\ref{A0}) being   a solution again, and for that  the  terms  which  are expressed in $ F $ and its derivatives, all be   equal to zero.  This means that the dominant dynamical equations over  the special solution  (\ref{A0}) are the same standard  equations (\ref{W1}) and (\ref{W2}). Therefore, we expect  $ F $ and all its derivatives  that appear in the above equations  would be zero simultaneously for the special solution (\ref{A0}). This goal is only possible if $ F $ is considered as a  function of the powers of three special  scalars; the scalars that all are zero for the special solution  (\ref{A0}). It is easy  to show that these scalars are:
\begin{eqnarray} \label{sc2}
  &&{\cal S}_{1}=\partial_{\mu}\theta\partial^{\mu}\theta-\omega_{s}^{2},\\&&
 {\cal S}_{2}=\partial_{\mu}R\partial^{\mu}R-R^6+R^4,\\&&
 {\cal S}_{3}=\partial_{\mu}R\partial^{\mu}\theta.
\end{eqnarray}
For example, if one considers  $F=F({\cal S}_{1}^{n},{\cal S}_{2}^{n},{\cal S}_{3}^{n})$, it leads to
\begin{eqnarray} \label{sc3}
 && \frac{\partial}{\partial x^{\mu}}\left(\frac{\partial F }{\partial (\partial_{\mu}R)}\right)=\sum_{i=1}^{3} \left[n(n-1){\cal S}_{i}^{(n-2)}\frac{\partial {\cal S}_{i}}{\partial x^{\mu}}\frac{\partial {\cal S}_{i} }{\partial (\partial_{\mu}R)} \frac{\partial F}{\partial Z_{i}}+n{\cal S}_{i}^{(n-1)}\frac{\partial}{\partial x^{\mu}}\left(\frac{\partial {\cal S}_{i} }{\partial (\partial_{\mu}R)} \frac{\partial F}{\partial Z_{i}}\right) \right] \nonumber  \\&&
\frac{\partial F}{\partial R}=\sum_{i=1}^{3}\left[n{\cal S}_{i}^{(n-1)}\frac{\partial {\cal S}_{i}}{\partial R}\frac{\partial F}{\partial Z_{i}}\right]\nonumber\\&&
\frac{\partial}{\partial x^{\mu}}\left(\frac{\partial F }{\partial (\partial_{\mu}\theta)}\right)=\sum_{i=1}^{3} \left[n(n-1){\cal S}_{i}^{(n-2)}\frac{\partial {\cal S}_{i}}{\partial x^{\mu}}\frac{\partial {\cal S}_{i} }{\partial (\partial_{\mu}\theta)} \frac{\partial F}{\partial Z_{i}}+n{\cal S}_{i}^{(n-1)}\frac{\partial}{\partial x^{\mu}}\left(\frac{\partial {\cal S}_{i} }{\partial (\partial_{\mu}\theta)} \frac{\partial F}{\partial Z_{i}}\right) \right] \nonumber,
\end{eqnarray}
where $ Z_ {i} = {\cal S}_{i}^n $. It is easy to understand that if $ n \geq 3 $, then all of these terms  would  be zero for the special solution (\ref{A0}). In general, it can be shown that the  all combinations according  to  the following series have the  desired wanted  features:
\begin{eqnarray} \label{sf0}
 F=\sum_{n_{3}=0}^{\infty}\sum_{n_{2}=0}^{\infty}\sum_{n_{1}=0}^{\infty} a({n_{1},n_{2},n_{3}}){\cal S}_{1}^{n_{1}}{\cal S}_{2}^{n_{2}}{\cal S}_{3}^{n_{3}},
\end{eqnarray}
provided $ (n_ {1} + n_ {2} + n_ {3}) \geq 3 $. Here,  coefficients $ a ({n_{1}, n_{2}, n_{3}})$ are some arbitrary functions of the allowed scalars (except $\theta$). Accordingly, the new  purposed  system (\ref{LN}) is essentially an extended nonlinear  Klien-Gordon system. In fact, it consists of two parts,   the original part ${\cal L}$  which is a standard KG system, and the additional term $F$,  which must be functions of the powers of the kinetic scalars.

 For  some special  choices   of the series  (\ref{sf0}), the stability of the  special  solution (\ref{A0}) is guaranteed  properly. For example, first let us  consider  three linear independent combinations of scalars $ {\cal S}_{1} $, $ {\cal S}_{2} $ and $ {\cal S}_{3} $ as follows:
 \begin{eqnarray} \label{e5}
  &&{\cal K}_{1}= R^2 {\cal S}_{1},\\&&
{\cal K}_{2}= R^2 {\cal S}_{1}+{\cal S}_{2}, \\&&
{\cal K}_{3}= R^2 {\cal S}_{1}+{\cal S}_{2}+2 R {\cal S}_{3}.
\end{eqnarray}
Then, the proper additional  functional  $ F $ can be introduced   as follow:
\begin{equation} \label{F}
 F=\sum_{i=1}^{3} A_{i}(
 {\cal K}_{i})^3,
 \end{equation}
In which $ A_{i} $'s ($i=1,2,3$) are only three large constants. However, for this particular choice (\ref{F}),  the corresponding energy density function is
\begin{eqnarray} \label{MTE}
&&\varepsilon(x,t)=\left[\dot{R}^2+R'^2+R^2(\dot{\theta}^2+\theta'^2)+V(R)\right]+\nonumber\\&&\quad\quad \quad \quad\sum_{i=1}^{3}\left[3A_{i}C_{i}
{\cal K}_{i}^{2}-A_{i}{\cal K}_{i}^3\right]=\varepsilon_{o}+\varepsilon_{1}+\varepsilon_{2}+\varepsilon_{3},
\end{eqnarray}
which  is divided  into four separate parts. Coefficients  $ C_ {i} $'s are
\begin{equation}\label{cof}
C_{i}=\dfrac{\partial{\cal K}_{i}}{\partial \dot{\theta}}\dot{\theta}+\dfrac{\partial{\cal K}_{i}}{\partial \dot{R}}\dot{R}=
\begin{cases}
\quad\quad 2 R^2 \dot{\theta}^{2} & \text{i=1}
\\
2( \dot{R}^{2}+ R^2 \dot{\theta}^2) & \text{i=2}
\\
2(\dot{R}+ R \dot{\theta})^{2}
 & \text{i=3}.
\end{cases}
 \end{equation}
 After a straightforward calculation,  given that $ \omega_{s}^2 = 100 $, one can obtain:
 \begin{eqnarray} \label{eis}
&&\varepsilon_{1}= {\cal K}_{1}^2 A_{1} [5 R^2\dot{\theta}^2+R^2\theta'^2+100 R^2]\geq 0,\\&&
\varepsilon_{2}={\cal K}_{2}^2 A_{2} [5 R^2 \dot{\theta}^2+5\dot{R}^2+ R^2 \theta'^2+U(R)]\geq 0, \\&&
\varepsilon_{3}={\cal K}_{3}^2 A_{3} [5( R \dot{\theta}+\dot{R})^2+( R \theta'+R')^2+U(R)]\geq 0,
\end{eqnarray}
in which
\begin{equation} \label{UR}
 U(R)= R^6-R^4+100 R^2.
 \end{equation}
This function (\ref{UR}) is always ascending and bounded from below by zero. Therefore, all terms in    $ \varepsilon_{1} $, $ \varepsilon_{2} $ and $ \varepsilon_{3} $ are  positive definite. For the special solution ($\ref{A0}$) and  vacuum state ($R=0$), all $ \varepsilon_{i} $'s  ($i=1,2,3$) would be zero simultaneously. It is reminded that $ {\cal K}_{i}$'s, like the scalars $ {\cal S}_{i} $'s, for the special solution  (\ref{A0}), all would be zero simultaneously. Since,   for the special solution (\ref{A0}) the energy contribution   belongs to the   additional term $F$ is practically zero (i.e. ${\cal K}_{i}=0$ then $\varepsilon_{F}=\sum_{i=1}^{3}\varepsilon_{i}=0$),  that is  why we call the additional term $F$ "\emph{phantom term}".

In general, since there are two independent scalar fields  $ R $ and  $ \theta $,  and depending on them three independent  scalars $ {\cal K}_{1} $, $ {\cal K}_{2} $ and  $ {\cal K}_{3} $ are considered, therefore,  except for the non-trivial solitary wave solution (\ref{A0}), it is not possible to find any other space-time functions $R(x,t)$ and $\theta(x,t)$  for which $ {\cal K}_{i} $'s      all being  zero simultaneously. Hence, for the other solutions of the new extended  system (\ref{LN} with \ref{F}), it is never possible for three independent scalars $ {\cal S}_{i}$'s or $ {\cal K}_{i} $'s  to be zero simultaneously. In other words, for the other solutions of the new system (\ref{LN} with \ref{F}), always at least one of the scalars $ {\cal K}_{i} $'s  is a nonzero function, and then  if constants $ A_{i} $'s are considered to be large numbers, at least one of the functions $ \varepsilon_{i} $'s ($i=1,2,3$)  becomes a non-zero large function, which implies that the energy of the other   solutions   would be always larger than the rest energy of the special  solution (\ref{A0}). Accordingly, the definition of the new extended  system (\ref{F}) for large values of $ A_ {i} $'s,  causes the special solution  (\ref{A0}) turns  to a  soliton solution with the minimum rest energy among the other solutions of the new extended system (\ref{F}). Note that, the first term of the energy density function ($ \varepsilon_ {o} $) has not   considered yet in the stability consideration. In the next section, we show that the contribution of this term (i.e. $\varepsilon_{o}$) is so small,  compared to the other terms $\varepsilon_{i}$'s, that essentially can be ignored  in the stability consideration, provided constants $A_{i}$'s are considered to be large numbers.

 \section{stability under small variations}\label{s5}

 In this section, we want to study the stability of the special  solution  (\ref{A0}) for the small deformations. In general, a little deformed  non-moving solitary wave solution  can be presented as follows:
 \begin{equation} \label{so1}
R(x,t)=R_{s}(x)+\delta R(x,t), \quad  \quad \theta(x,t)=\theta_{s}(t)+\delta \theta=\omega_{s}t+\delta \theta(x,t),
\end{equation}
where $R_{s}(x)=1/\sqrt{1+x^2}$ and $\theta_{s}=\omega_{s}t$, and $ \delta R $ and $ \delta\theta $ are any permissible   small space-time functions. Now, by inserting this deformed solution (\ref{so1}) in the first part of the energy density $ \varepsilon_{o} $ (\ref{MTE}) and keeping the terms in order of  $ \delta R $ and $ \delta \theta $, then it yields:
\begin{eqnarray} \label{so3}
&&\varepsilon_{o}(x,t)=\varepsilon_{os}(x)+\delta\varepsilon_{o}(x,t)\approx \left[ R_{s}'^2+R_{s}^2\omega_{s}^2+V(R_{s})\right]+\nonumber\\&&
\quad\quad\quad 2\left[R_{s}'(\delta R')+R_{s}(\delta R)\omega_{s}^2+
   R_{s}^{2}\omega_{s}(\delta\dot{\theta})+\frac{1}{2}\frac{dV(R_{s})}{dR_{s}}(\delta R)\right].
\end{eqnarray}
Note that for the special non-moving solution  (\ref{A0}), $ \dot{R_{s}} = 0 $, $ \theta_{s}' = 0 $ and $ \dot{\theta_{s}} = \omega_{s} $. It is clear that $ \delta \varepsilon_{o} $ is not necessarily an absolute positive function. Hence,  it may take negative values at some space-time points for some permissible variations $\delta R$ and $\delta\theta$.   Now, we do this for the other energy density parts  (\ref{MTE}):
 \begin{eqnarray} \label{so4}
&&\varepsilon_{i}(x,t)=\varepsilon_{is}+\delta\varepsilon_{i}=\delta\varepsilon_{i}=[3A_{i}(C_{is}+\delta C_{i})({\cal K}_{is}+\delta{\cal K}_{i})^{2}-A_{i}({\cal K}_{is}+\delta{\cal K}_{i})^{3}]=\nonumber\\&&
[3A_{i}(C_{is}+\delta C_{i})(\delta{\cal K}_{i})^{2}-A_{i}(\delta{\cal K}_{i})^{3}]\approx[3A_{i}C_{is}(\delta{\cal K}_{i})^{2}-A_{i}(\delta{\cal K}_{i})^{3}]\approx[6A_{i}\omega_{s}^2(\delta{\cal K}_{i})^{2}]\geqslant 0\nonumber\\&&
\end{eqnarray}
 Note that, for the special solitary  wave solution  (\ref{A0}):  $ \varepsilon_{is} = 0 $, $ {\cal K}_{is} = 0 $ and $ C_{is} = 2 \omega_{s}^2 $. As we expected, according to the previous equation (\ref{so4}),  $ \delta\varepsilon_{i} $'s ($i=1,2,3$) all are always positive definite.

The variation of the total energy density function ($\delta \varepsilon$) would  be equal to the sum of the changes in the all four separate parts, namely $ \delta \varepsilon = \delta \varepsilon_{o} + \sum_{i = 1}^{3} \delta \varepsilon_{i} $. Since $ \delta \varepsilon_{o} $ is not necessarily positive and $ \delta \varepsilon_{i} $'s are always positive,  therefore, the difference between  the order of magnitudes  of  $\sum_{i=1}^{3} \delta \varepsilon_{i} $ and $ \delta \varepsilon_{o} $ is important for various small deformations (\ref{so1}). In general, it can be easily shown that $ \delta{\cal K}_{i} $'s and $ \delta C_{i} $'s are in  the  order of the   first power of the  variations  $ \delta R $ and $ \delta \theta $. Therefore,  according to the result (\ref{so4}), $ \delta \varepsilon_{i} $'s   are in the order of $A_{i} ({\delta R})^{2}$,  $A_{i} (\delta \theta)^{2} $ and $A_{i}(\delta R \delta \theta)$,    while $ \delta \varepsilon_{o} $,  as mentioned above (\ref{so3}), is only in  the order of   $ \delta R $ and $  \delta \theta $.

On this basis, if  constants $A_{i} $'s are considered to be large numbers, it can be easily understood that the special solution (\ref{A0}) effectively leads  to a stable solution under the influence of the small variations. In fact, there is approximately an small  certain amount for the  order of the magnitude of the small variations $ \delta R $ and  $ \delta \theta$ in such a way that if the maximum of  them be larger than this  certain amount, $\delta\varepsilon$ would be positive and then the special solution is called a stable object for such variations. For example, if we assume a system with $ A_ {i} = 10^{40} $, then only  for the variations that  $ \textrm{max}\{O(\delta R), O ( \delta \theta)\}< 10^{- 20} $, the special  solution (\ref {A0}) dose  not have the   condition of  the stability. In other words, for these very small variations (i.e. $ \textrm{max}\{O(\delta R), O ( \delta \theta)\}< 10^{- 20} $), the total energy density variation ($ \delta \varepsilon $) may take  negative values.  But note that, these variations  are so small  which  physically can be ignored  in the stability considerations. However, if the order of the magnitude  of one of  the variations $\delta R$  and $ \delta \theta$ be larger than  $10^{-20}$ (i.e.  $\textrm{max}\{O(\delta R), O ( \delta \theta)\}> 10^{-20} $), then $\delta\varepsilon$ would be positive and the stability is  guaranteed appreciably for such small variations.

 Since the special solution (\ref{A0}) does not have the stability  condition for very small  variations, it  is a sign  of the fact that the dominant dynamical equation for the special solution (\ref{A0}) is the same standard  nonlinear KG equations (\ref{W1}) and (\ref{W2}). In other words, for very small unimportant  variations, the condition of stability  is violated (i.e. $ | \delta\epsilon_{o}| \geq \sum_{i = 1}^{3} \delta \varepsilon_{i} $), and the role of terms  which  containing   functional  $ F $, in comparison with other terms,  would be ineffective.

Through a biologic example, we try to make the story better. Understanding the unstable nature of the special  solution  (\ref{A0}) for very low variations, less than the certain amount, is like  a chicken   in the  egg. Chicken  inside the egg can slightly  move  its head, hands and feet, and it does not necessarily have a permanent  fixed status. But, allowed movements for chicken within egg, due to egg shell, are subject to severe constraints, for example, it can not  send out their feet through  the egg shell. In other words, regardless  some unimportant movements inside the egg, due to the solid egg shell,  the chicken has to be  in the same shape of the    egg.  For this simulation, firstly, the phantom term  $ F $ has the role of the  egg shell, of course a massless and very solid shell, i.e. the  phantom  term  $ F $ requires that the particle shape (\ref{A0}) always stays in a specified format; secondly, the chicken  itself  is similar to the same special solution (\ref{A0}); and thirdly, the internal permissible movements  of the chicken inside the egg are  like  some permissible so small variations of the special solution (\ref{A0}) for which $\delta\varepsilon<0$. Briefly, the new phantom term (\ref{F}) in the new extended system (\ref{LN}) behaves like a stability catalyzer and does  not have any role in the other observable such as rest mass and charge.

Some one may think that this model is an artificial model which is properly tuned to satisfied  all wanted features  specially for a purposed solitary wave  solution  (\ref{A0}).  But note that,              firstly; many of the models in the standard quantum field  theory are artificial, i.e. many people  try to tune   them    manually  so that  lead to  desired results and expectations, secondly; it  have not been yet a classical relativistic field model which  leads to an energetically   stable non-topological soliton solution and this one is the first, thirdly;  the required  features  which we introduced in the beginning of the section \ref{s3} for the special particle-like function (\ref{A0}), forced us to reach this strange model, i.e. it is unique   and we could  not find another  completely different   one which satisfied the all wanted  features properly.

 \section{conclusion}\label{s6}

 Considering the above points, it is theoretically possible to speak of a special  non-topological  soliton solution (\ref{A0}) for which the dominant dynamical equation would be a special kind of the standard nonlinear KG equations. This was done if one  add a new proper scalar term ($F$) to the original KG Lagrangian density (\ref{cnk}). This new term behaves like a massless  phantom which  surrounds the particle-like solution  (\ref{A0}) and resists any significant changes in its internal structure. For a free solitary wave solution  (\ref{A0})  or one whose  internal structure is deformed  very small, the dominant dynamical equations are the  same standard nonlinear KG equations (\ref{W1}) and (\ref{W2}), and the contribution  of the phantom term $ F $ in its dominant dynamical equations is zero or negligible. In other words, just for this  special  solution (\ref{A0}), the complicated  dynamical  equations  (\ref{geq}) and (\ref{geq2}) reduce to the same  standard nonlinear  KG equations (\ref{W1}) and (\ref{W2}) respectively.
But for some significant changes that are larger than a certain amount, the actual role of the added term   $ F $ is significant and resists  such changes. In other words, the presence of such changes in the internal structure of the particle need  high  external energies. The power and ability of this new term $ F $ for  the stability of the special  solution  (\ref{A0}) is related to three constants $ A_ {i} $ ($ i = 1,2,3$)  so that the larger values of those  make the phantom term stronger. The role of the new (phantom) term $ F $ for the special solution (\ref{A0}) just makes it stable, but does not affect its other physical properties, such as mass, charge, shape and the  specific  dominant dynamical  equations, i.e. it acts like a stability catalyzer.

 In general, the new system is one of the extended nonlinear KG systems that leads to very complicated dynamical  equations. In general, such systems are thought to be nonstandard and distant in the standard  theory of classical and quantum  fields, but for the case presented in this paper, i.e. Eqs.~(\ref{LN}) and (\ref{F}), its solutions have energies larger  than the rest energy of the special  solitary wave solution  (\ref{A0}). In other words, the special solitary wave  solution  (\ref{A0}) is truly a non-topological energetically stable  soliton solution.

\end{document}